\documentclass[namedreferences,hyperref,optionalrh]{spr-sola}
\usepackage{graphicx}        % For eps figures, newer & more powerfull
\usepackage{amssymb}         % useful mathematical symbols
\usepackage{color}           % For color text: \color command
\usepackage{natbib}
\usepackage{tabularx}
\usepackage{gensymb}
\usepackage[margin=1in]{geometry}

\chardef\us=`\_

\begin{document}

\begin{frontmatter}

\title{High-resolution Observation of Blowout Jets Regulated by Sunspot Rotation}

\author[addressref={aff1,aff10},email={tygou@ustc.edu.cn; tingyu.gou@cfa.harvard.edu}]{\inits{T.}\fnm{Tingyu}~\snm{Gou}\orcid{0000-0003-0510-3175}}
\author[addressref=aff1,email={rliu@ustc.edu.cn}]{\inits{R.}\fnm{Rui}~\snm{Liu}\orcid{0000-0003-4618-4979}}
\author[addressref=aff2]{\inits{Y.}\fnm{Yang}~\snm{Su}\orcid{0000-0002-4241-9921}}
\author[addressref=aff3]{\inits{A.}\fnm{Astrid}~\lnm{M.}~\snm{Veronig}\orcid{0000-0003-2073-002X}}
\author[addressref=aff1]{\inits{H.}\fnm{Hanya}~\snm{Pan}\orcid{0000-0002-0318-8251}}
\author[addressref=aff1]{\inits{R.B.}\fnm{Runbin}~\snm{Luo}\orcid{0000-0002-2559-1295}}
\author[addressref={aff2,aff4}]{\inits{W.}\fnm{Weiqun}~\snm{Gan}}

\address[id=aff1]{CAS Key Laboratory of Geospace Environment, University of Science and Technology of China, Hefei 230026, China}
\address[id=aff2]{Key Laboratory of Dark Matter and Space Astronomy, Purple Mountain Observatory, Chinese Academy of Sciences, Nanjing 210023, China}
\address[id=aff3]{Institute of Physics \& Kanzelh\"{o}he Observatory for Solar and Environmental Research, University of Graz, 8010 Graz, Austria}
\address[id=aff4]{University of Chinese Academy of Sciences, Nanjing 211135, China}

\address[id=aff10]{T.G. has moved to Center for Astrophysics $|$ Harvard \& Smithsonian, 60 Garden Street, Cambridge, MA 02138, USA}

\runningauthor{T. Gou et al.}
\runningtitle{High-resolution Observation of Blowout Jets}

\begin{abstract}
Coronal jets are believed to be the miniature version of large-scale solar eruptions. In particular, the eruption of a mini-filament inside the base arch is suggested to be the trigger and even driver of blowout jets. Here we propose an alternative triggering mechanism, based on high-resolution H$\alpha$ observations of a blowout jet associated with a mini-filament and an M1.2-class flare. The mini-filament remains largely stationary during the blowout jet, except that it is straddled by flare loops connecting two flare ribbons, indicating that the magnetic arcade embedding the mini-filament has been torn into two parts, with the upper part escaping with the blowout jet. In the wake of the flare, the southern end of the mini-filament fans out like neighboring fibrils, indicative of mass and field exchanges between the mini-filament and the fibrils. The blowout jet is preceded by a standard jet. With H$\alpha$ fibrils moving toward the single-strand spire in a sweeping fashion, the standard jet transitions to the blowout jet. The similar pattern of standard-to-blowout jet transition occurs in an earlier C-class flare before the mini-filament forms. The spiraling morphology and sweeping direction of these fibrils are suggestive of their footpoints being dragged by the leading sunspot that undergoes clockwise rotation for over two days. Soon after the sunspot rotation reaches a peak angular speed as fast as 10 deg~hr$^{-1}$, the dormant active region becomes flare-productive, and the mini-filament forms through the interaction of moving magnetic features from the rotating sunspot with satellite spots/pores. Hence, we suggest that the sunspot rotation plays a key role in building up free energy for flares and jets and in triggering blowout jets by inducing sweeping motions of fibrils.
\end{abstract}

\keywords{Flares; Coronal jets; Filaments; Magnetic fields; X-ray emission}

\end{frontmatter}
%-------------------------------------------------

\section{Introduction} \label{sec:intro}

Jets are ubiquitous in the solar atmosphere, characterized by a collimated ejection of plasma and an inverted-Y morphology with a base arch and a spire \cite[]{Raouafi2016,Shen2021}. Coronal jets are often dichotomized into standard jets and blowout jets \cite[]{Moore2010,Sterling2015}. The latter is different from the former in that the base arch does not remain static and closed but explodes open, hence is considered as a miniature version of the large-scale eruptions that are associated with coronal mass ejections (CMEs). A mini-filament is often found inside the base arch of coronal jets, and apparently the failed or successful eruption of the mini-filament corresponds to the standard- or blowout-jet morphology, respectively \cite[]{Sterling2015}. Similar to their large-scale counterpart \cite[]{Wang2000}, mini-filaments may serve as a marker of sheared or twisted magnetic field \cite[]{Mackay2010}. Hence it has been argued that the dichotomy of coronal jets might result from ``the dichotomy of base arches that do not have and base arches that do have enough shear and twist to erupt open'' \cite[]{Moore2010}. If this is true, there must exist a critical threshold of magnetic shear or magnetic twist, or magnetic helicity, to determine if a base arch is capable of eruption or not. The existence of such a threshold remains contentious for active regions \cite[]{Toriumi&Wang2019}, and its existence for coronal jets is also elusive. 

Nevertheless, it is important to understand how the magnetic free energy of coronal jets is built up through the evolution of the photosphere. It has been noticed that coronal jets are often occurring in regions in which satellite spots continually collide with the main sunspot, with the presence of coronal null points, or in regions where magnetic elements of opposite polarities cancel each other through converging and/or shearing motions \cite[]{Shen2021}. 

On the other hand, nature seldom agrees with a simple dichotomy. For example, there is a continuum spectrum of filament eruptions \cite[]{Gilbert2007}, ranging from failed eruptions in which both the filament mass and the associated magnetic structure are confined in the lower corona, to partial eruptions in which either the filament mass \cite[e.g.,][]{Tripathi2009} or the associated magnetic structure partly escape \cite[e.g.,][]{Liu2007}, and to complete eruptions, not to mention similarly diverse eruptive behaviours of double-decker filaments \cite[see a review in][]{Liu2020}. It raises questions as to whether there also exists a continuum spectrum of mini-filament eruptions, and how their eruptive behaviors affect the associated jets. 

Below we present high-resolution observations of a mini-filament on 2022 November 11 using the 1-m New Vacuum Solar Telescope \cite[NVST;][]{LiuZh2014} in China, and investigate its evolution in association with the sunspot rotation in the NOAA active region (AR) 13141 and the role of sunspot rotation in a series of jets that occurred in the same AR.

\section{Data and Instruments}

In this study, we use both the line-of-sight and vector magnetograms observed by the Helioseismic and Magnetic Imager \citep[HMI;][]{Scherrer2012} onboard the Solar Dynamics Observatory \citep[SDO;][]{Pesnell2012}. The line-of-sight magnetograms are obtained every 45~s with a pixel scale of 0.6$''$. The vector magnetograms for space-weather HMI active-region patch (SHARP) are produced every 12~min and remapped with the cylindrical equal area (CEA) projection. With HMI vector magnetograms, we examine the injection rate of magnetic helicty [Mx$^2$ s$^{-1}$] across the photosphere in the active region, and use the Differential Affine Velocity Estimator for Vector Magnetograms \citep[DAVE4VM;][]{Schuck2008} to obtain photospheric velocities. We further calculated the relative helicity flux across the photospheric boundary $S$ \cite[]{Berger1984}:
\begin{equation}
\left.\frac{dH}{dt}\right|_S=2\int_S(\mathbf{A}_p\cdot\mathbf{B}_t)V_{\perp n}\,dS-2\int_S(\mathbf{A}_p\cdot\mathbf{V}_{\perp t})B_n\,dS, \label{eq:helicity}
\end{equation}
where $\mathbf{A}_p$ is the vector potential of the reference potential field $\mathbf{B}_p$; $t$ and $n$ refer to the tangential and normal directions, respectively. Obtained from DAVE4VM, $\mathbf{V}_\perp$ is the photospheric velocity that is perpendicular to magnetic field \cite[]{Liu&Schuck2012}. The first term on the right of the equation describes the emergence of twisted flux tubes into the corona; the second term describes photospheric motions that shear and braid magnetic field lines.

The flares in AR 13141 are observed by the Atmospheric Imaging Assembly \citep[AIA;][]{Lemen2012} onboard SDO and by NVST in H$\alpha$. AIA takes full-disk images of the Sun with a pixel resolution of 0.6$''$ and a temporal cadence of 12~s in seven EUV channels. We mainly use images in 131~\AA\ and 304~\AA\ passbands to study the flare activities and jets in AR 13141. The NVST H$\alpha$ images are taken from 01:40 to 08:40~UT on 2022 November 11, and we use the linecenter (6562.81~\AA) images with a temporal cadence of 72~s and a pixel scale of 0.165$''$.

The hard X-ray (HXR) emissions from the flares are observed by the Hard X-ray Imager \citep[HXI;][]{Zhang2019,Su2019} onboard the Advanced Space-based Solar Observatory \citep[ASO-S;][]{Gan2019} and by the Spectrometer Telescope for Imaging X-rays \citep[STIX;][]{Krucker2020} onboard Solar Orbiter \citep{Muller2020}, both of which provide HXR imaging of the Sun using a Fourier-transform indirect-imaging technique but observing from different locations with a separation angle of $\sim22\degree$ on November 11, 2022. HXI images the full solar disk in the energy range of $\sim$10--300~keV \citep{Su2024} with a spatial resolution as high as 3.1$''$. The flares studied in this work were observed during the early stage of HXI and the HXR images are not available below 20 keV. HXR images by HXI are reconstructed using HXI\_Clean method with the preliminary calibration parameters and the sub-collimator groups G3-G10, which result in a spatial resolution of $\sim6.5''$. When on its Sun synchronous orbit ASO-S passes through the South Atlantic Anomaly region or the radiation belt, we use the HXR data from STIX, which provides HXR spectroscopy in the 4--150 keV energy range. On 2022 November 11, Solar Orbiter was separated from the Earth by about $-22\degree$ in longitude, $+4.5\degree$ in latitude, at a heliocentric distance of 0.62~AU from the Sun. The observation times of STIX light curves and images are corrected to account for the time difference of light traveling to Earth and Solar Orbiter. We reconstruct the STIX HXR images in the energy ranges of 6--10~keV, 15--25~keV, and 25--45~keV (if there is any enhanced emission) with a time integration of 1~min. The HXR images are primarily reconstructed using the CLEAN algorithm, while MEM\_GE and EM algorithms \citep{Massa2019,Massa2020} are also performed for further consistency check especially for the HXR sources in the high-energy energy band. The STIX images are re-projected to the Earth's perspective and are overlaid on AIA and NVST images to study the flare X-ray emissions. 

\section{Analyses and Results}

\begin{figure}[htbp]
	\centering
	\includegraphics[width=\textwidth]{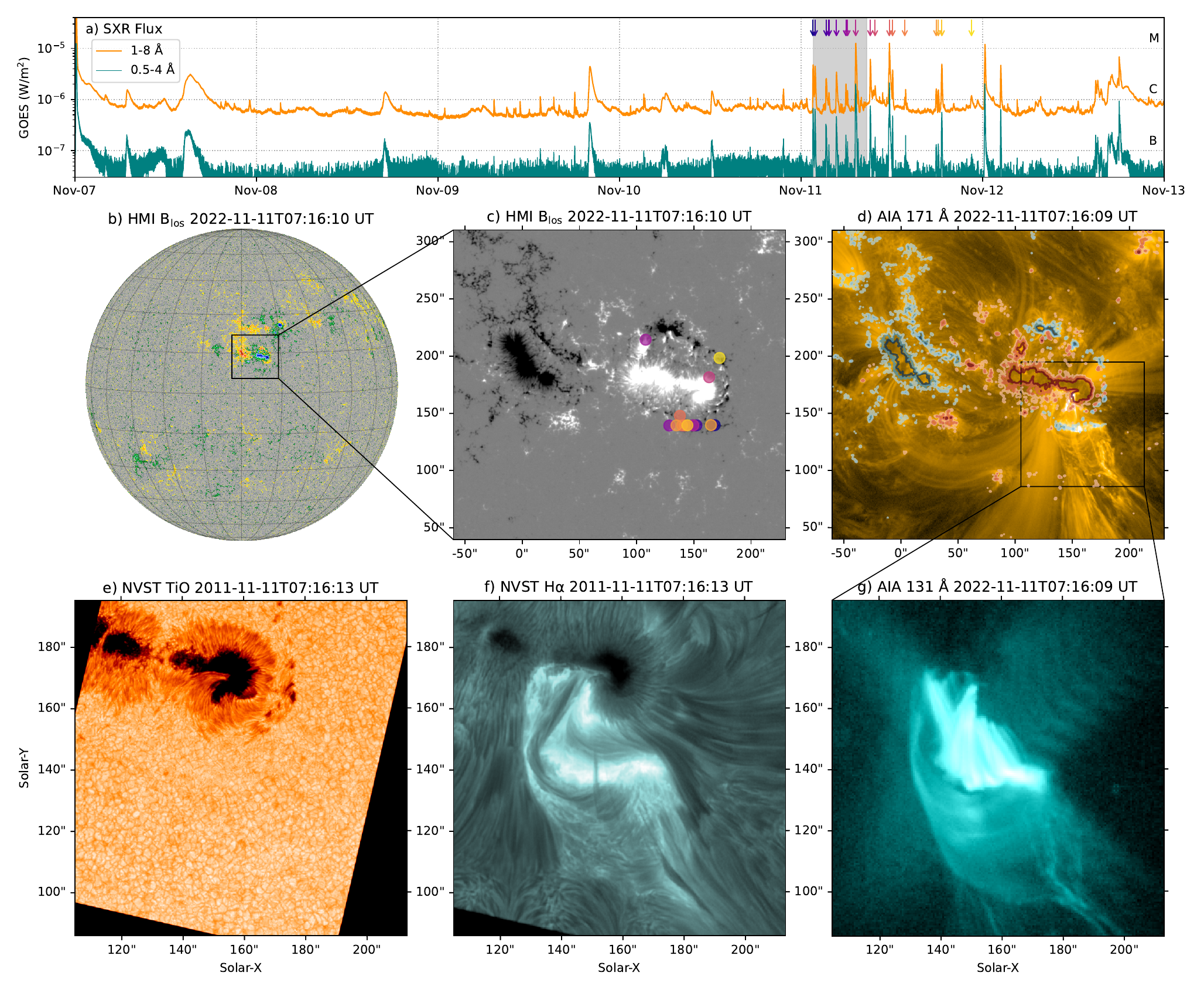}
	\caption{\small Overview of solar activities in AR 13141. a) GOES 1--8~\AA\ (orange) and 0.5--4~\AA\ (green) SXR fluxes. The colored arrows mark the peak time of flares on 2022 November 11 in Table~\ref{tab:list}. The light gray shade indicates the observation period of NVST. b) SDO/HMI full-disk line-of-sight magnetogram on 2022 November 11. The black rectangle indicates the FOV of panels (c,d). c) Zoom-in view of AR 13141. The colored dots mark the locations (after the correction of solar rotation) of flares on November 11 as listed in Table~\ref{tab:list}, whose peak times are marked by arrows with the same colors in panel (a). d) SDO/AIA 171~\AA\ image overplotted with HMI B$_{\rm los}$ contours at $\pm$300, $\pm$600, and $\pm$1000 Gauss. The black rectangle indicates the FOV of panels (e--g), which show NVST TiO and H$\alpha$ as well as AIA 131~\AA\ images, respectively.}
	\label{fig:overview}
\end{figure}

\begin{table}
    \centering
    \caption{\small List of flares in AR 13141 on 2022 November 11.}
    \begin{tabularx}{\textwidth}{@{\extracolsep{\fill}}ccccc}
        \hline
        No. & Start Time & Peak Time & GOES Class & Location\\
        \hline
        1 & 01:30 & 01:35 & C6.8 & N11W07\\
        2 & 01:44 & 01:50 & C6.5 & N11W07\\
        3 & 03:10 & 03:20 & C2.5 & N11W07\\
        4 & 03:36 & 03:41 & C1.3 & N11W07\\
        5 & 04:33 & 04:41 & C3.3 & N11W07\\
        6 & 05:46 & 05:56 & C1.9 & N11W07\\
        7 & 06:04 & 06:09 & C1.3 & N16W06\\
        8 & 07:00 & 07:14 & M1.2 & N11W09\\
        9 & 08:58 & 09:09 & C5.9 & N14W11\\
        10& 09:44 & 09:47 & C1.7 & N11W10\\
        11& 11:27 & 11:40 & M1.2 & N11W11\\
        12& 12:02 & 12:06 & C3.6 & N12W11\\
        13& 13:35 & 13:43 & C1.0 & N11W12\\
        14& 17:45 & 17:52 & C1.5 & N11W14\\
        15& 18:03 & 18:09 & C1.4 & N11W16\\
        16& 18:22 & 18:37 & C4.7 & N11W15\\
        17& 22:21 & 22:32 & C1.1 & N15W19\\
        \hline
    \end{tabularx}
    \label{tab:list}
\end{table}

\subsection{Overview of AR 13141}
AR 13141 shows up at the solar northeast limb on 2022 November 4 and rotates to the backside of the Sun after 2022 November 17. The AR produces a series of C-class and a few M-class flares during its pass over the visible solar disk, which makes it the dominant flaring region on the Sun (Figure~\ref{fig:overview}). Especially on 2022 November 11, it produces 17 flares of GOES C-class and above, including two moderate M-class flares (Table~\ref{tab:list}); on November 12, it produces 10 more flares, including an M1.1 one; while it is relatively dormant during earlier days (e.g., November 7--10), as shown by the GOES soft X-ray (SXR) fluxes (Figure~\ref{fig:overview}a). All of these flares occurred in the western part of the AR, mostly between the main spots with positive polarities and satellite spots with negative polarities (Figure~\ref{fig:overview}c).

\begin{figure}[htbp]
	\centering
	\includegraphics[width=\textwidth]{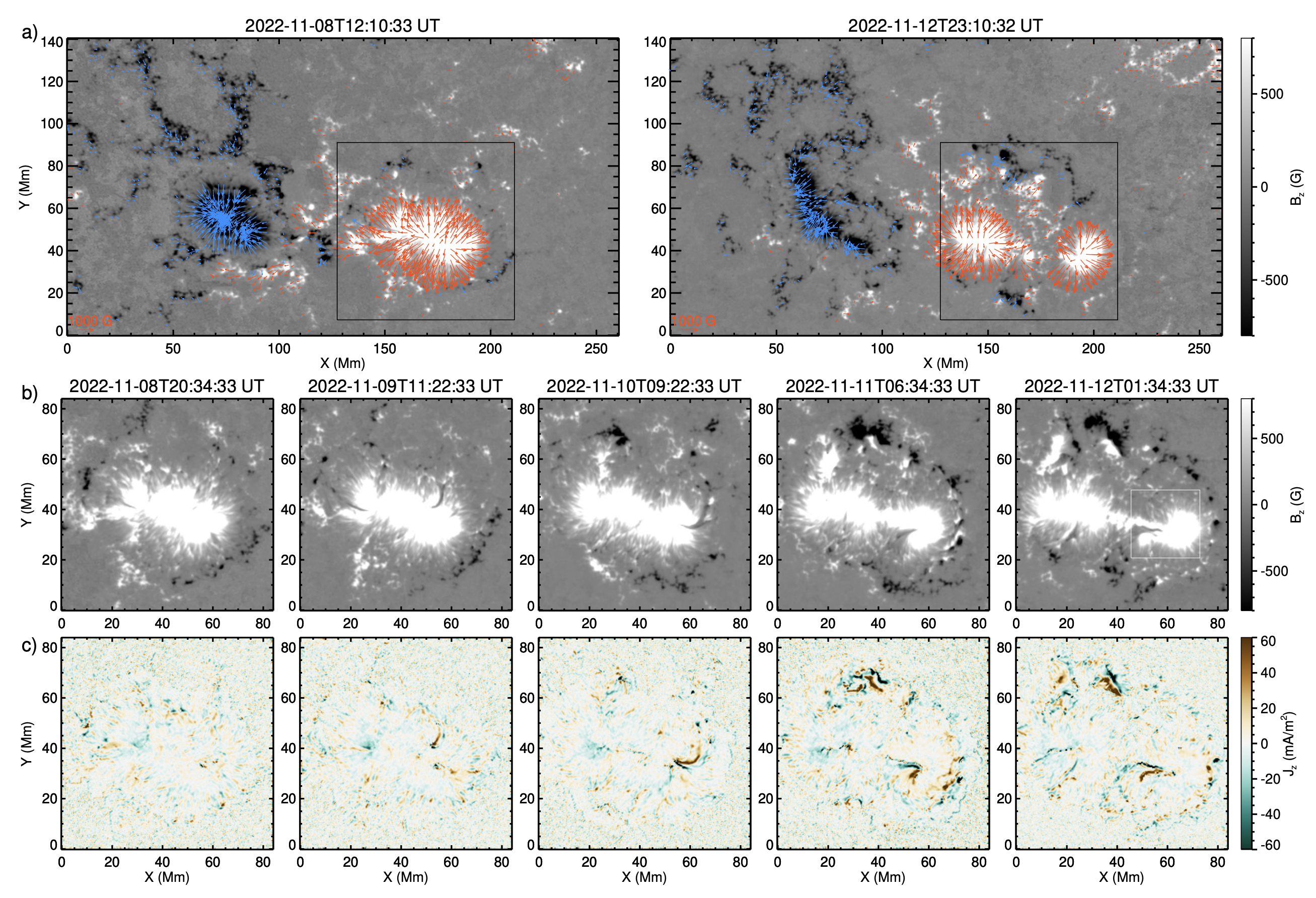}
	\caption{\small Evolution of magnetic field and current density. a) Vector magnetograms of AR 13141 on November 8 and 12. The FOV in panel (a) is selected for the DAVE4VM calculation. The black rectangles indicate the FOV of the bottom panels (b,c). Orange and blue arrows show the tangential magnetic field components in the positive and negative polarities, respectively. b,c) Vertical component of magnetic fields and current densities at different times. The white rectangle in the middle right panel indicates the FOV plotted in Figure~\ref{fig:dave}. An animation of this figure is available.}
	\label{fig:bvec_jz}
\end{figure}

\begin{figure}[htbp]
	\centering
	\includegraphics[width=\textwidth]{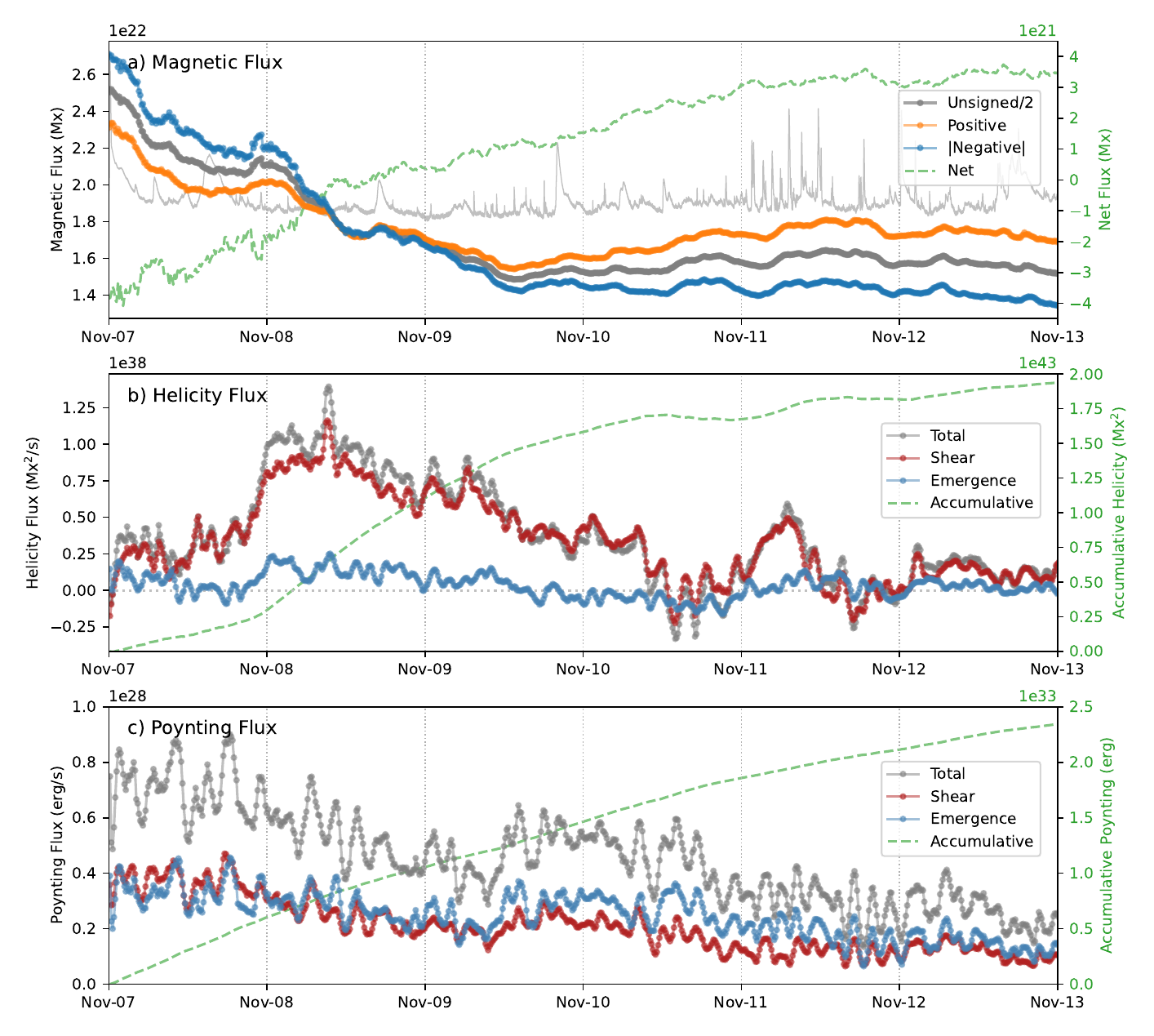}
	\caption{\small Long-term evolution of AR 13141. a) Magnetic flux within the FOV in Figure~\ref{fig:bvec_jz}a, scaled by the left y-axis and shown in orange (positive flux), blue (negative flux in magnitude), and dark gray (unsigned flux divided by 2); net flux is shown as dashed green curve and scaled by the right y-axis. The GOES 1--8~\AA\ SXR flux (light gray) are plotted in an arbitrary unit. b) Magnetic helicity injection in AR 13141. The total (gray), shear term (red), and emergence term (blue) of helicity injection rates are scaled by the left y-axis. The accumulative helicity (dashed green) is scaled by the right y-axis. c) Same as those in panel (b) but for the Poynting fluxes.}
	\label{fig:helicity_flux}
\end{figure}

The AR consists of a leading sunspot with positive polarity and a trailing one with negative polarity as it rotates toward the solar disk center (Figure~\ref{fig:bvec_jz}a). Both become elongated and transform into complex sunspot groups as time progresses. In particular, the leading spot eventually separates into two major spots, surrounded by satellite spots and pores with negative polarities (Figure~\ref{fig:bvec_jz}b). 

Overall, the AR's evolution is characterized by flux cancellation until mid November 9 (Fig.~\ref{fig:helicity_flux}a) and injection of positive helicity [Mx$^2$~s$^{-1}$] on November 8 and 9 (Fig.~\ref{fig:helicity_flux}b), which is dominated by the plasma flows, i.e., the shear term of helicity injection \cite[2nd term in Eq.~\ref{eq:helicity};][]{Liu&Schuck2012}.

\subsection{Sunspot Rotation and Separation}

\begin{figure}[htbp]
	\centering
	\includegraphics[width=\textwidth]{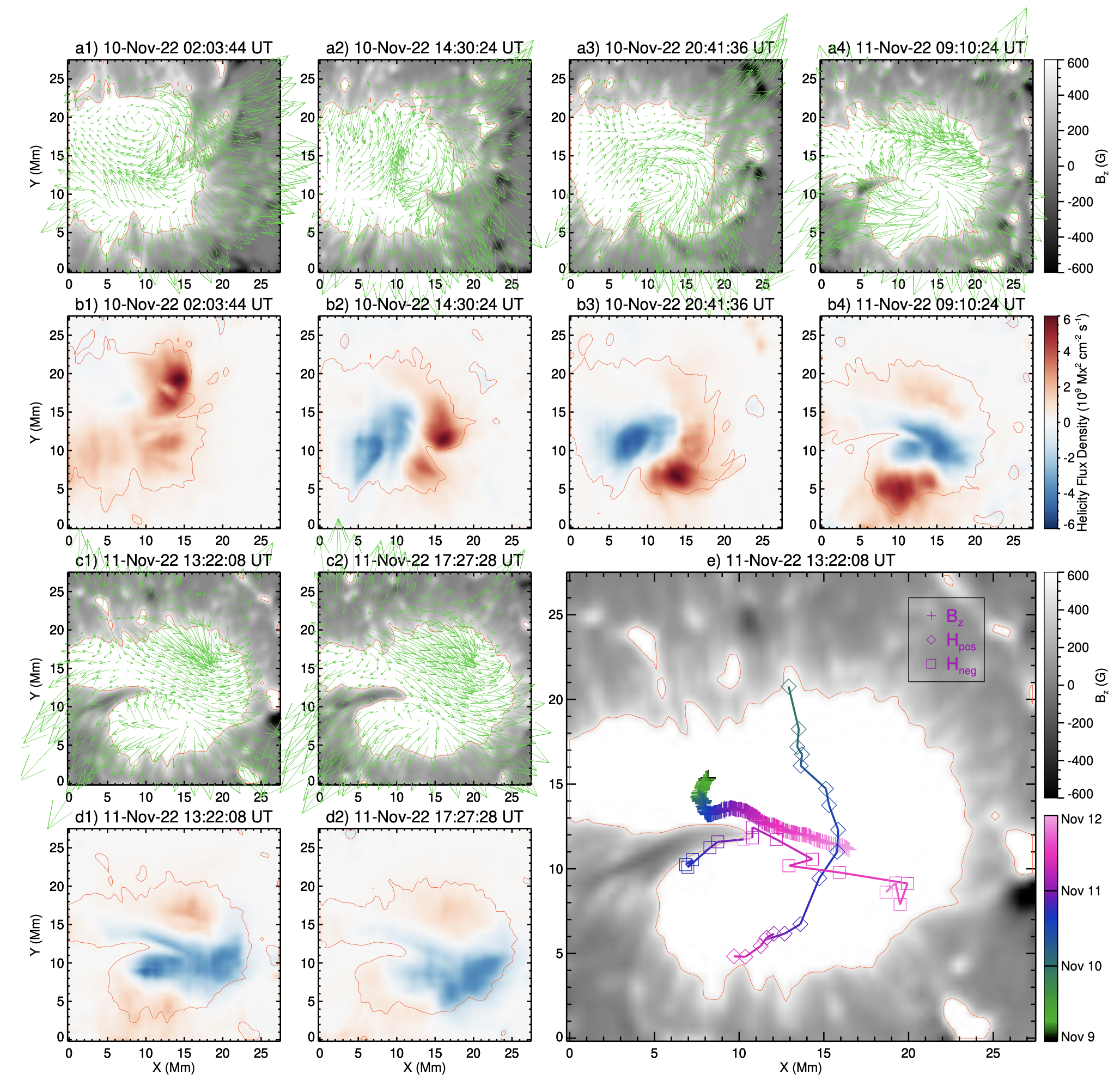}
	\caption{\small Evolution of DAVE flow fields and injection of magnetic helicity flux densities associated with the sunspot. The results are averaged over 2 hrs around the time of each map. The orange contours outline locations where B$_{\rm z}$~=~500~G, which serves as an indication of the sunspot. The green arrows in panels (a,c) indicate tangential DAVE velocities, which are overlaid on the closest HMI B$_{\rm z}$ maps. The red and blue colors in the helicity flux density maps in panels (b,d) represent the positive and negative values of injected helicities (only shear term plotted in the figure). Panel (e) shows the temporal evolution of identified centroid locations during Nov 9--12, overlaid on a B$_{\rm z}$ map on Nov 11, coded by colors in two color bars in the right, respectively. The plus symbols indicate the centroid locations of the sunspot weighted by normal magnetic fluxes of B$_{\rm z} \geqslant$ 500~G. The diamond and square symbols indicate the centriod locations weighted by injections of positive and negative helicity fluxes, respectively.}
	\label{fig:dave}
\end{figure}

\begin{figure}[htbp]
	\centering
	\includegraphics[width=\textwidth]{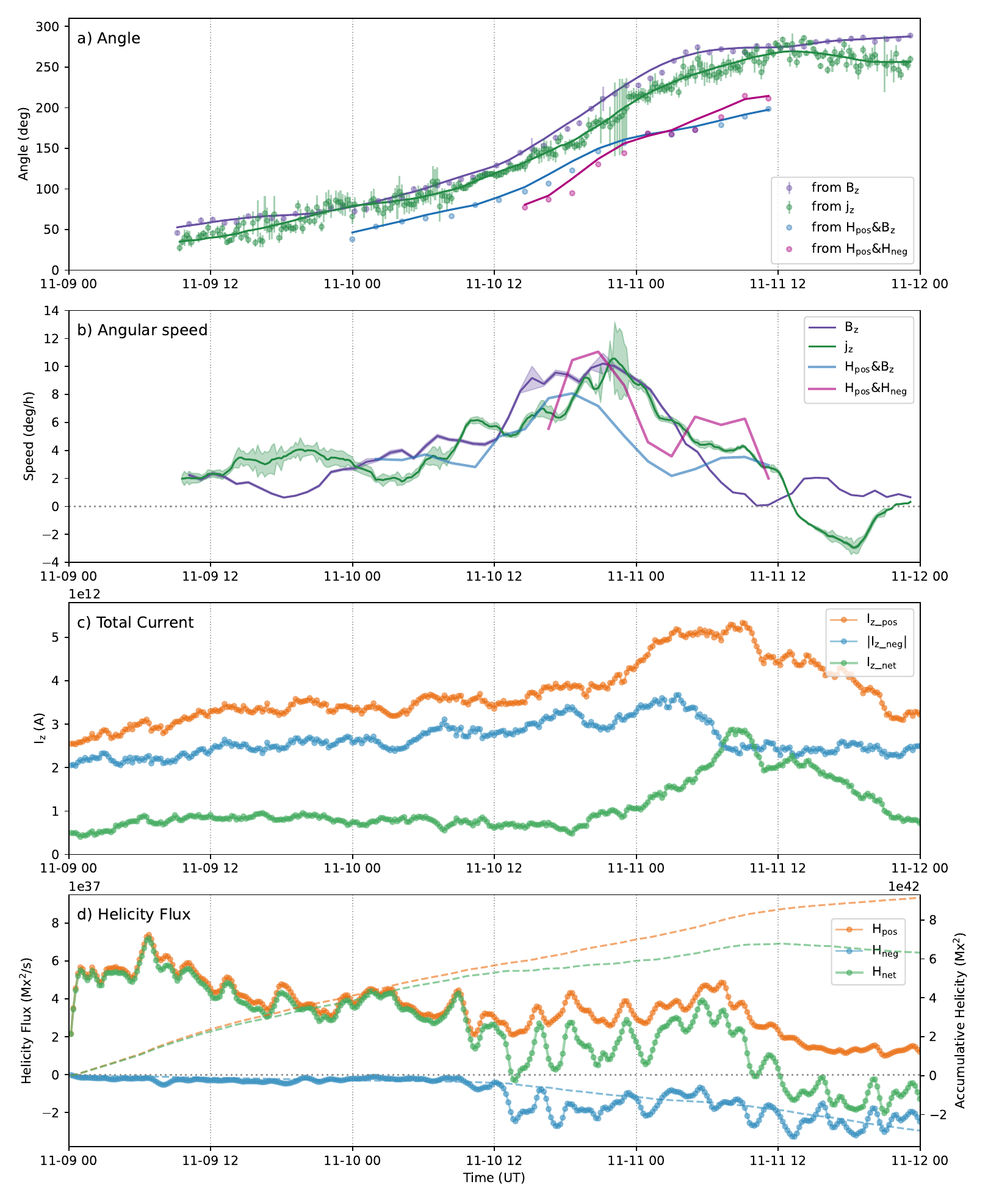}
	\caption{\small Evolution of the sunspot. a) Rotation angles with respect to the solar north, measured from the patterns of B$_{\rm z}$, j$_{\rm z}$, and from the centroids of B$_{\rm z}$ and positive helicity flux, centroids of positive and negative helicity fluxes, respectively (see the text for details). The colored solid curves are the corresponding smoothed results after applying the Savitzky--Golay smoothing filter. b) Angular speeds of the sunspot rotation derived from the smoothed angles in panel (a). c) Total vertical currents inside the sunspot region. d) Injection of helicity fluxes (shear term only) in the sunspot. The temporal injection rates (dotted lines) are scaled by the left y-axis, and the accumulative helicity flux injections (dashed curves) are scaled by the right y-axis.}
	\label{fig:sunspot_param}
\end{figure}

The westernmost sunspot in the leading sunspot group rotates clockwise as it separates from the rest of sunspot group. The rotation is visible not only in the maps of B$_z$ but also in the maps of $J_z$, the vertical component of electric current density (see Figure~\ref{fig:bvec_jz}b,c and its accompanying animation). The rotation is observed to start at around 9 UT on November 9, when the sunspot is notched in the northwestern quadrant by weak magnetic fields (gray colors) in the B$_{\rm z}$ map. This notch serves as a reference to mark the clockwise rotation of the sunspot (Figure~\ref{fig:bvec_jz}b). A pair of current ribbons with opposite signs but large magnitude $\geqslant$~60~mA/m$^{2}$ is found to make a similar V shape as the notch, and together they rotate with the sunspot (Figure~\ref{fig:bvec_jz}c). As shown by the DAVE method, in the sunspot region the typical rotating flow speed is about 0.3 km/s (Figure~\ref{fig:dave}; averaged over 2 hrs), which, however, does not represent the overall speed of the sunspot rotation. 

The sunspot is originally dominated by injection of positive magnetic helicity flux [Mx$^2$ cm$^{-2}$ s$^{-1}$] before November 10 (Figures~\ref{fig:dave}b1\&\ref{fig:sunspot_param}d). Starting from around 12:00~UT on November 10, negative helicities are injected into the sunspot (Figure~\ref{fig:dave}b2). Eventually, it is the negative helicity injection that begins to dominate in the sunspot by about 13~UT on November 11 (Figure~\ref{fig:dave}d1). We identify the centroid locations of the positive and negative helicity flux injections inside the sunspot (diamond and square symbols in Figure~\ref{fig:dave}e), weighted by H$_{\rm pos}\geqslant 3\times 10^{9}$~Mx$^2$~cm$^2$~s$^{-1}$, H$_{\rm neg}\leqslant -2.5\times 10^{9}$~Mx$^2$~cm$^2$~s$^{-1}$, respectively. Associated with the sunspot motion, the centroid of positive helicity fluxes moves from north to south (diamond symbols) and that of negative helicity fluxes from east to west (square symbols) as demonstrated in Figure~\ref{fig:dave}e. This is consistent with our impression that the patches of enhanced helicity fluxes of both signs seem to rotate about each other, along with the rotation of the sunspot notch, especially during the late November 10 and early November 11 (Figure~\ref{fig:dave}a,b).

To quantify the sunspot rotation, we measure the rotation angles (with respect to the solar north) of different features. First, we measure the rotation angle of the sunspot notch in B$_{\rm z}$ maps, by linearly fitting four manually-selected points along the southern edge of the notch as it appears (only the central straight part is used as it curves outward). We take B$_{\rm z}$ maps with 1-hr cadence for the manual measurement and repeat the measurement three times to obtain the average value and measurement errors. The results are plotted as purple dots with error bars in Figure~\ref{fig:sunspot_param}a. Second, we measure the rotation of the current ribbons in $J_z$ maps (Figure~\ref{fig:bvec_jz}c). We use the contour of j$_{\rm z}$~=~40~mA/m$^2$ to obtain the pattern of the positive current ribbon. Although the current ribbon is overall curved, its inner 2/3 portion close to the center of the sunspot is almost straight, and this section is used for linear fitting to obtain the angle. The results are plotted with green dots in Figure~\ref{fig:sunspot_param}a. Since the positive current ribbon is observed to lag behind the rotating notch in B$_{\rm z}$ (Figure~\ref{fig:bvec_jz} and its animation), the measured angles are generally smaller than those measured from B$_{\rm z}$ maps. In addition, we measure the angle subtended by the centriods of magnetic flux and positive helicity flux (plus and diamond symbols, respectively, in Figure~\ref{fig:dave}e), and the angle subtended by the centroids of positive and negative helicity fluxes (diamond and square symbols, respectively, in Figure~\ref{fig:dave}e), and the results are shown as light blue and magenta dots in Figure~\ref{fig:sunspot_param}a, respectively. 

To derive the rotating speed of the sunspot, we first smooth the measured data points with the Savitzky–Golay method (solid lines in Figure~\ref{fig:sunspot_param}a). The angular rotation speeds derived from the four different datasets are generally consistent with each other (Figure~\ref{fig:sunspot_param}b), showing a long-duration rising phase lasting about 36 hours, which reaches a peak value of about 10$\degree\,\mathrm{hr}^{-1}$ at the end of November 10, and a following decay phase of about 12 hours until the speed decreases to zero at about 12~UT on November 11. The sunspot rotates over 240$\degree$ within two days. Both the total rotational angle and the transient angular speed of the sunspot are larger than many previously reported results \citep[e.g.,][]{Brown2003,Zhang2007,Yan2007,Torok2013,Bi2016}. By identifying the sunspot centroids weighted by the magnetic flux (B$_{\rm z} \geqslant$ 500~G; colored plus symbols in Figure~\ref{fig:dave}e), one can see that the sunspot starts to move westward at about 12~UT on November 10, which roughly coincides with a steep rise in the angular rotation speed.

We obtain the total vertical electric currents and the rate of helicity injection (shear term) [Mx$^2$~s$^{-1}$] inside the rotating sunspot (within the $B_z=500$~G contour in Figure~\ref{fig:dave}) and study the temporal evolution (Figure~\ref{fig:sunspot_param}c,d). Both the positive and negative currents increase with time as the sunspot rotates fast (Figure~\ref{fig:sunspot_param}c). On November 11 after the sunspot's angular speed peaks, the net current flowing through the sunspot reaches a higher level than the preceding days (the green curve in Figure~\ref{fig:sunspot_param}c). For the helicity injection by shearing motions, the sunspot is dominated by positive helicity injection in the beginning (Figure~\ref{fig:sunspot_param}d). However, from about 12~UT on November 10, negative helicities start to inject into the sunspot (see also helicity density maps in Figure~\ref{fig:dave}b), which coincides with the steep increase in the angular rotation speed (Figure~\ref{fig:sunspot_param}b) and the westward motion (Figure~\ref{fig:dave}e) of the sunspot. The sunspot is dominated by negative helicity injection from about 12~UT on November 11 onward (see the green dotted curve in Figure~\ref{fig:sunspot_param}d), when the sunspot rotation is almost halted (Figure~\ref{fig:sunspot_param}b).

\subsection{Flare Activities}

\begin{figure}[htbp]
	\centering
	\includegraphics[width=\textwidth]{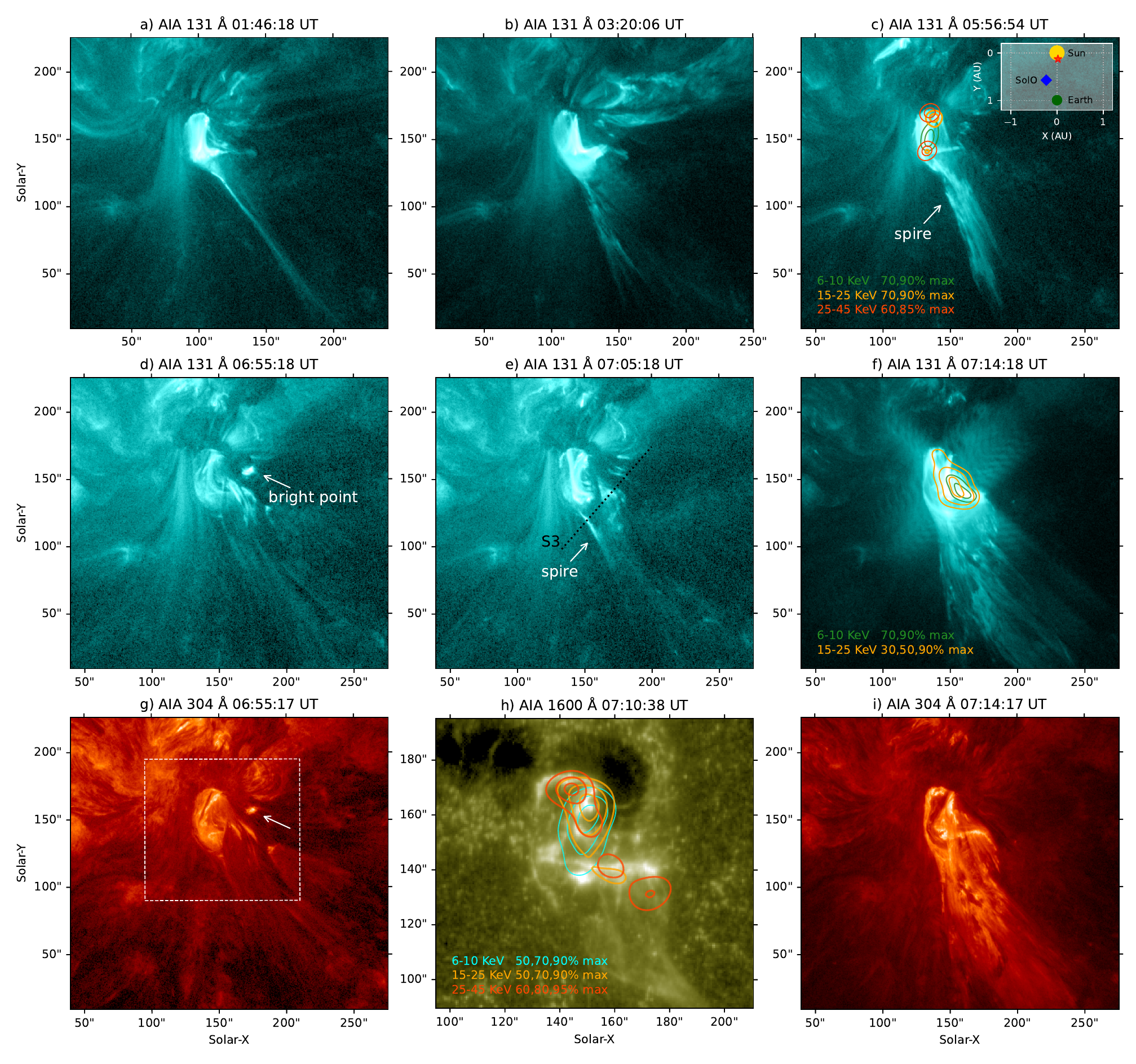}
	\caption{\small SDO/AIA observations of flares on 2022 November 11. Panels (a--c) show the C-class flares \#2, \#3, and \#6 listed in Table~\ref{tab:list}, respectively. Panels (d--i) show the M1.2 flare (\#8). The inserted plot in panel (c) shows the location of Solar Orbiter (SolO) on November 11 relative to the Sun and Earth. The red star symbol indicates the location of flares on the Sun. The STIX HXR sources in different energy bands obtained using the CLEAN algorithm are overlaid on AIA images in panels (c,f,h). The black dotted line in panel (e) indicate the virtual slit S3 to generate the stack plot in Figure~\ref{fig:slit}e. The white dotted rectangle in the panel (g) indicate the FOV of panel (h) and the NVST H$\alpha$ images in Figure~\ref{fig:halpha}. An animation of this figure is available.}
	\label{fig:flares}
\end{figure}

\begin{figure}[htbp]
	\centering
	\includegraphics[width=\textwidth]{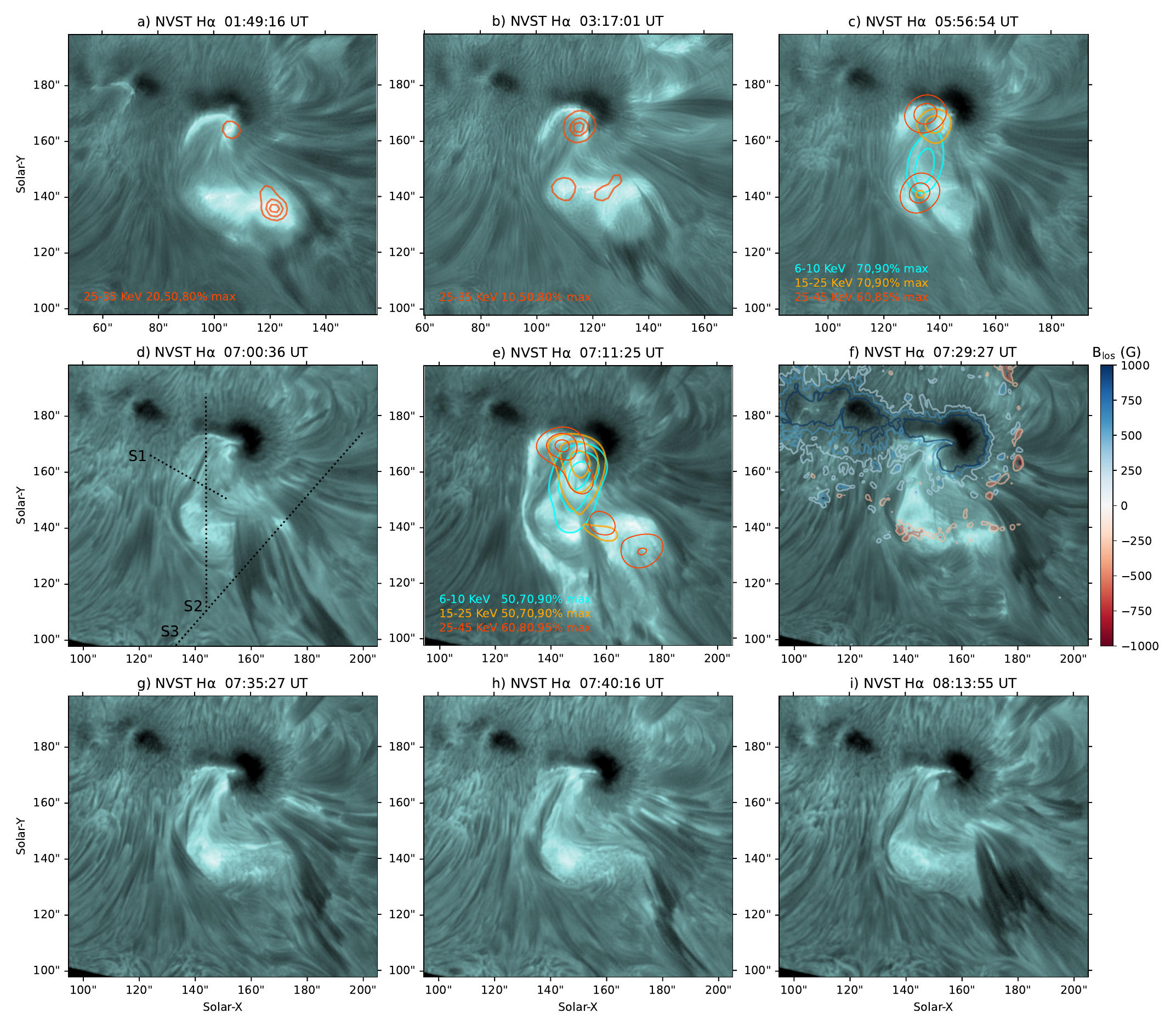}
	\caption{\small Snapshots of NVST H$\alpha$ observation on 2022 November 11. The HXR sources reconstructed using the CLEAN algorithm from the HXI observation are overlaid in panels (a,b). The STIX HXR sources are overlaid in panels (c,e), which are the same as those in Figure~\ref{fig:flares}(c,h). The dotted lines in panel (d) indicate the virtual slits used to generate the stack plots in Figure~\ref{fig:slit}, where labels S1--S3 mark the origins of slits. The colored contours in panel (f) show the simultaneous HMI magnetogram at B$_{\rm los}$ = $\pm$300, $\pm$600, and $\pm$1000 Gauss. An animation of this figure is available.}
	\label{fig:halpha}
\end{figure}

\begin{figure}
    \centering
    \includegraphics[width=\textwidth]{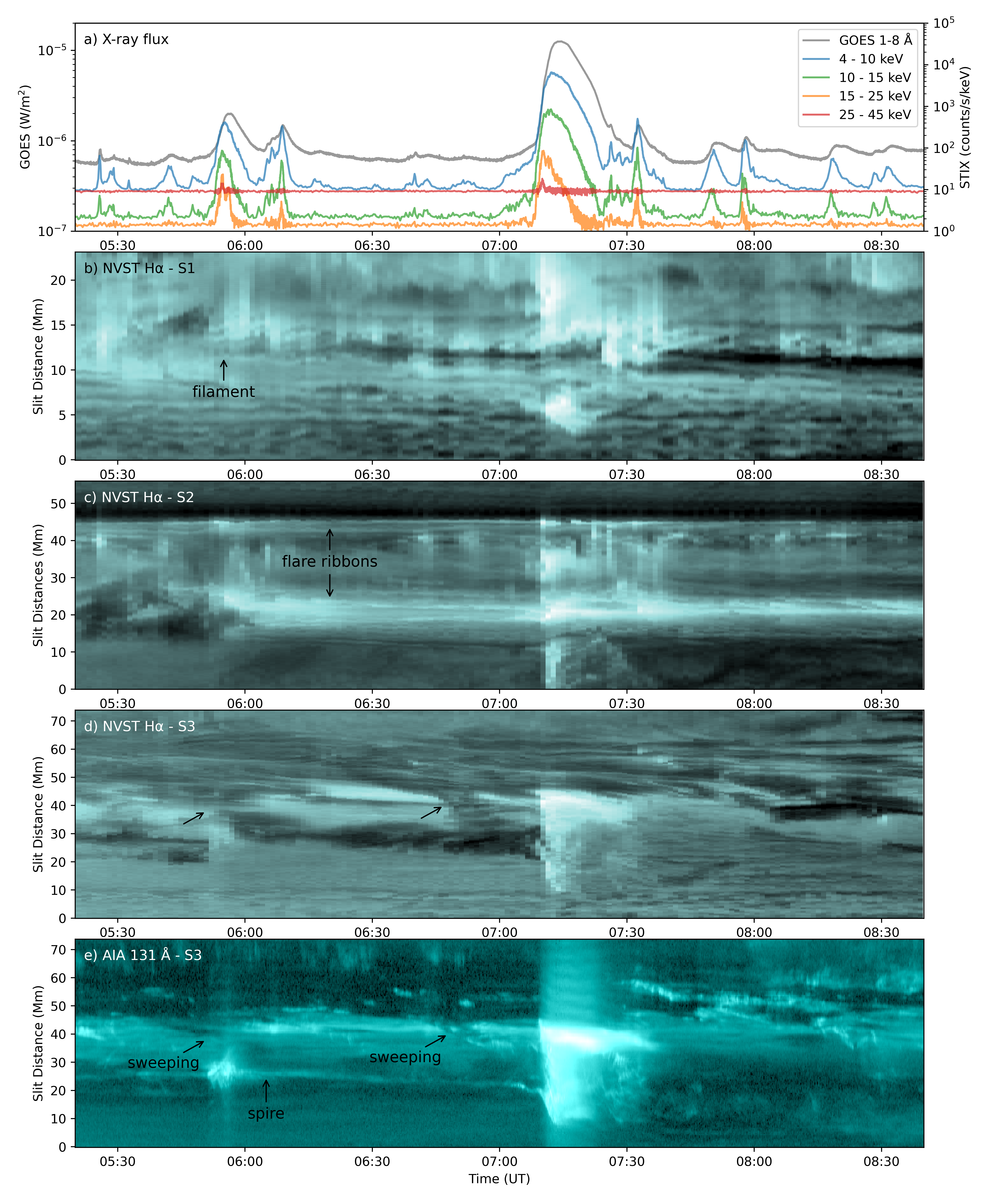}
    \caption{Temporal evolution of the mini-filament and flares. a) GOES 1--8~\AA\ SXR flux (gray; scaled by the left y-axis) and STIX HXR light curves (colored lines, scaled by the right y-axis). b--e) Stack plots of slits S1--S3 in Figures~\ref{fig:halpha}(d)\&\ref{fig:flares}(e). In panels (d) and (e), arrows mark the sweeping motions of fibrils preceding two blowout jets at 05:50 and 07:10 UT, respectively. } 
    \label{fig:slit}
\end{figure}

A series of flares occur in AR 13141 (e.g., Figure~\ref{fig:flares} and its animation), after the sunspot rotation reaches its peak angular speed and starts to move westward. Most flares originate from the peripheral regions of the rotating sunspot (Figure~\ref{fig:overview}c; except for Flare \#7 in Table~\ref{tab:list} in the northeast) and show similar morphologies, often associated with a jet (Figure~\ref{fig:flares}). Bright flare loops form at the base of coronal jets and connect to two flare ribbons, where HXR footpoint sources with energies $>$25~keV are observed (Figs.~\ref{fig:flares}\&\ref{fig:halpha}). The northern HXR footpoint is located near the sunspot penumbra associated with strong magnetic field strengths, but does not show a significant asymmetry in HXR fluxes compared with the other footpoint (Figure~\ref{fig:halpha}), which is the usual case in many other flares \citep[e.g.,][]{Yang2012}. The HXR emissions from the footpoints are similar to the cases in \citet{Saqri2024}, which includes a microflare in the same AR on November 10 showing unusually hard energy spectra and efficient particle acceleration. In between the two footpoints there is a third source at lower X-ray energies, which is interpreted as the coronal HXR source located at the top of the flare loops (e.g., Fig.~\ref{fig:flares}c,f,h). Particularly, flares \#6 and \#8 are each associated with a blowout jet. The M1.2-class flare (\#8) well conforms with the classic, two-ribbon-flare morphology, with both ribbons visible in AIA 1600~{\AA} (Fig.~\ref{fig:flares}h). The ribbon associated with positive polarities is partially located in the sunspot; the other ribbon associated with negative polarities is located to the south of the rotating sunspot. Flare loops are observed to connect the two ribbons (Fig.~\ref{fig:flares}f).

High-resolution NVST H$\alpha$ observations show a mini-filament connecting the sunspot and the southern negative-polarity patches (Figs.~\ref{fig:overview}f\&\ref{fig:halpha} and the accompanying animation). The northern footpoint of the min-filament seems to be rooted in around the tip of the notch of the rotating sunspot (Figs.~\ref{fig:overview}e,f\&\ref{fig:halpha}f), where the injection of negative magnetic helicities becomes dominant since about 12 UT on November 10 (Figs.~\ref{fig:dave} \& \ref{fig:sunspot_param}d), the day before the flares. We put a virtual slit (S1 in Fig.~\ref{fig:halpha}d) to cut across the northern portion of the mini-filament to investigate its dynamic evolution (see the stack plot in Fig.~\ref{fig:slit}b). In H$\alpha$, the mini-filament becomes prominent after about 6 UT on November 11 (Figure~\ref{fig:slit}b), with threads apparently twisting about each other in a left-handed sense, in the wake of a C-class flare (\#6; see the accompanying animation of Fig.~\ref{fig:halpha}). During the subsequent M1.2-class flare (\#8), the filament becomes thicker and darker (Figure~\ref{fig:slit}b), and the left-handed twist becomes more evident (e.g., Fig.~\ref{fig:halpha}g and its animation).

Flare \#6 at 05:56 UT is associated with a small blowout jet featured by a curtain-like spire (Fig.\ref{fig:flares}c and its animation), but in the wake of the blowout jet a single-strand spire remains and persists until the early phase of the M1.2-class flare (Flare \#8; Fig.~\ref{fig:flares}d \& e). Hence the M-class flare initiates as a standard jet, which is featured by the single-strand spire located side by side with a thick bundle of fibrils (Figs.~\ref{fig:flares}g and \ref{fig:halpha}d) and a miniature flare-arcade bright point located to the west of the base arch (e.g., at 06:55~UT, Fig.~\ref{fig:flares}(d \& g)). Inside the base arch, a mini-filament is clearly seen in 304~{\AA}. By 07:05~UT, while the bright point has dimmed, multiple bright loops across the mini-filament appear in 131~{\AA} (Fig.~\ref{fig:flares}e and its animation), indicating that the magnetic fields overlying the mini-filament have started to participate in magnetic reconnection. Meanwhile, the dark, spiraling H$\alpha$ fibrils located to the west of the filament move clockwise (especially after 06:30 UT) to approach the filament in a sweeping manner (see the animation of Figure~\ref{fig:halpha}). We adopt two different slits (S2 and S3 in Fig.~\ref{fig:halpha}d) to show this dynamic evolution in H$\alpha$. Slit S2 is made to cut across the two flare ribbons. Slit S3 is made to cut across both the dark fibrils and the jet spire. In Fig.~\ref{fig:slit}d, one can see that starting from about 06:45 UT, the thick bundle of H$\alpha$ fibrils (slit position $\sim\,$30 Mm) immediately neighboring to the single-strand jet spire become darker and thicker, while farther away fibrils (slit position $\sim\,$35-50 Mm) move fast toward the spire, as indicated by the steepening of dark strips in the stack plot, which is also associated with the further darkening of the filament (Fig.~\ref{fig:slit}b). At about 07:10 UT, around the peak of HXRs at 15--25 and 25--45 keV, the thick bundle of fibrils suddenly disappear, and in AIA 304~{\AA} it is replaced by a bright curtain-like spire, along which materials spray outward (see the animation accompanying Fig.~\ref{fig:halpha}). These features are consistent with the occurrence of a blowout jet. The earlier, smaller blowout jet associated with Flare \#6 is similarly preceded by an episode of clockwise fibril sweeping at about 05:50 UT. Both episodes of fibril motions are also registered in the stack plot made from AIA 131~{\AA} images (Fig.~\ref{fig:slit}e).

With the onset of the blowout jet at 07:10 UT, two flare ribbons appear in H$\alpha$ on both sides of the mini-filament (Fig.~\ref{fig:slit}c), with flare loops connecting the two ribbons in AIA 131~{\AA}. These features indicate that the magnetic structure embedding the mini-filament also participates in the blowout jet. However, the mini-filament does not erupt as in typical blowout jets. Instead, it remains largely stationary, except that its southern end rooted in the negative polarity region become less well defined than before (Fig.~\ref{fig:halpha}e--i and its animation). At about 07:35 UT, the southern portion of the filament seems to fan out like neighboring fibrils (Fig.~\ref{fig:halpha}g), indicating that interchange reconnections might be ongoing between the filament field and the fibril field.

\subsection{Magnetic Configuration of the Flaring region} \label{subsec:magnetic_config}

To investigate the magnetic structure associated with the mini-filament and its evolution during the M-class flare, we studied the magnetic field configuration of the flaring region by extrapolating coronal magnetic fields, using HMI SHARP data as the boundary condition. These vector magnetograms are ``pre-processed'' to best match the force-free condition, before being fed into the ``weighted optimization'' non-linear force-free field (NLFFF) code \citep{Wiegelmann2006}. Here NLFFF is constructed over a uniform grid of $640\times 360\times 360$ pixels (pixel size 0.36 Mm), covering the major magnetic elements of AR13141 (cf. Fig.~\ref{fig:overview}c), and magnetic connectivities are investigated by calculating the map of squashing factor $Q$ \citep{Titov2002}. We distinguished open from close field lines. The former have one end at the upper or side boundaries of the box region that is selected for the NLFFF extrapolation, i.e., they are locally open; the latter have both ends at the bottom of the box region. We also obtained the maps of twist number $T_w=\int \nabla\times\mathbf{B}\cdot\mathbf{B}/4\pi B^2\,ds$, where the line integration is performed along each individual field lines \citep{Liu2016}.  

\begin{figure}[t]
	\centering
	\includegraphics[width=0.8\textwidth]{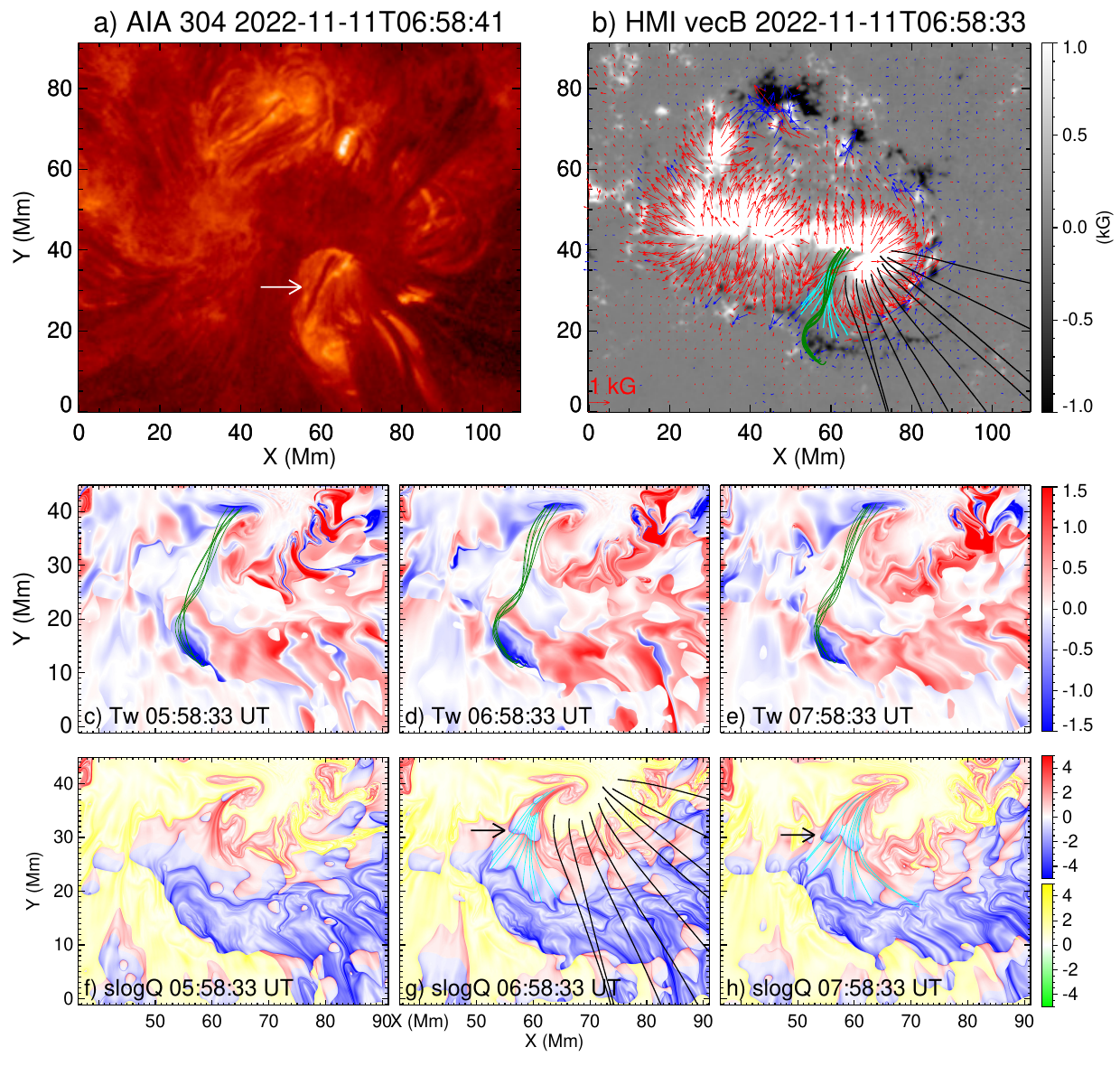}
	\caption{\small Magnetic configuration related to the M1.2-class flare on 2022 November 11. a) AIA 304~{\AA} image taken just before the flare onset, which is remapped with the CEA projection to match the HMI SHARP Vector magnetogram (b) taken at the same time. The arrow in (a) marks the filament of interest. In (b) the background image shows $B_z$ as scaled by $\pm1000$ G; overplotted vectors are the transverse field component originating from positive (red) and negative (blue) polarities. (c--e) maps of twist number $T_w$, where representative twisted field lines traced from $T_w$-enhanced regions are shown in dark green. (f--h) maps of signed squashing factor, $\mathrm{slog}\,Q= \mathrm{sign(B_z})\times \log Q$, where representative field lines that graze the surface at the bald patch segment (marked by the arrows in (g \& h)) are shown in cyan. The green-yellow and blue-red color tables indicate open- and close-field regions, respectively. In open-field regions, the green (yellow) color indicates negative (positive) polarities, while in close-field region the blue (red) color indicates negative (positive) polarities. In (g) a few locally open field lines that have similar orientations as the fibrils undergoing sweeping motions are shown in black. The field lines in (d) and (g) are reproduced in (b).}
	\label{fig:bp}
\end{figure}

Using the $T_w$ maps, we identify a magnetic flux rope (MFR) co-located with the mini-filament, by tracing magnetic field lines from $T_w$-enhanced regions (dark green curves in Fig.~\ref{fig:bp}). This MFR is formed as early as $\sim$02~UT on November 11, coinciding with the formation of a filament channel, where the chromospheric fibrils are roughly aligned along the PIL (Fig.~\ref{fig:halpha}a,b). Soon afterwards, the mini-filament is formed in the channel as evidenced in high-resolution NVST H$\alpha$ images (Fig.~\ref{fig:halpha} and its animation). At about 07~UT a bald-patch \cite[BP;][]{Titov1993} segment appears below the MFR in the $Q$-map (Fig.~\ref{fig:bp}g, marked by a black arrow). Representative field lines that are tangential to the surface at this BP segment are shown in cyan colors. However, both the MFR and the BP survives the M-class flare that peaks at 07:14 UT (Flare \#8; Fig.~\ref{fig:bp}(e \& h)), without exhibiting any obvious changes. 

In addition, from the maps of signed squashing factor, $\mathrm{slog}\,Q= \mathrm{sign(B_z})\times \log Q$ (Fig.~\ref{fig:bp}(f--h)), one can see that the fibrils undergoing sweeping motions are likely rooted in the open-field regions surrounding the northern end of the mini-filament, where magnetic field lines have similar footpoint locations and projected orientations as the fibrils. However, one must keep in mind that the cold materials of fibrils may not show the full length but only segments of the field lines. On the other hand, these open field lines are not associated with a null point or a bald-patch separatrix surface, as revealed in some jets originating from the edges of active regions \cite[e.g.,][]{Schmieder2013,Chandra2017,Wyper2019}. This could be partly attributed to the magnetic complexity of the active region under investigation as well as to the limit posed by the NLFFF modeling.

\section{Conclusion and Discussion}
\begin{figure}
    \centering
    \includegraphics[width=\textwidth]{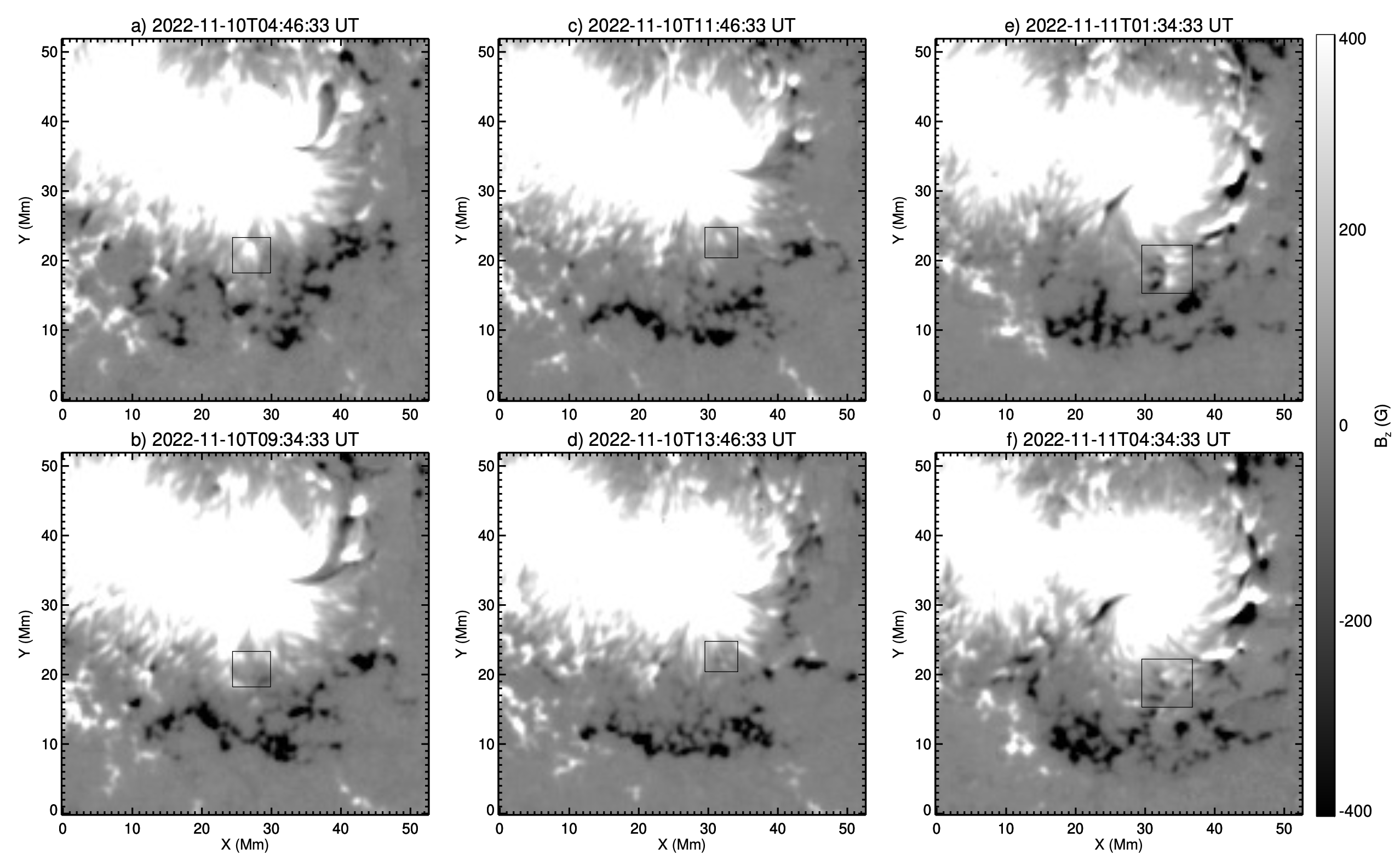}
    \caption{\small Evolution of HMI magnetograms around the flare region. Panels (a,b) (c,d), and (e,f) show three examples of the evolution of moving mangetic features originating from the rotating sunspot, respectively, as indicated by the black rectangles.} 
    \label{fig:bz}
\end{figure}

The energy buildup for coronal jets in active regions is often associated with a satellite spot \cite[e.g.,][]{Sakaue2017}. In our case, the sunspot rotation appears to be the dominant driving force of the active-region evolution. Since a sunspot is a monolithic region of concentrated magnetic flux, any rotation would act to twist up a large amount of flux. With the twist being transported into the corona via Alfv\'{e}n waves, the active region field would hold more free energy to be released in the form of jets, flares, and CMEs \cite[]{Brown2003,Tian2008}. As expected, the active region becomes flare-productive soon after the angular rotation speed of the sunspot peaks by the end of Nov 10 (Fig.~\ref{fig:sunspot_param}b), and most of the flares occur in the periphery of the rotating sunspot (Fig.~\ref{fig:overview}c). In particular, the mini-filament channel associated with the blowout jet and M1.2-class flare (\#8) is developed where moving magnetic features originating from the rotating sunspot continually advance into the satellite spots/pores with negative polarities and are subsequently canceled therein (Fig.~\ref{fig:bz}). 

The evolution of the mini-filament during the M1.2-class flare could be easily misidentified as a failed eruption. Only in high-resolution observations is the mini-filament revealed to be interchanging mass and field with the neighboring fibrils, which is evidenced by the fanning-out of its southern end, exhibiting similar morphological features as the neighboring fibrils (Fig.~\ref{fig:halpha}g). However, since it is largely stationary, the mini-filament cannot be the driver the blowout jet, as envisaged in some cartoon models \cite[e.g.,][]{Sterling2015}. Nevertheless, the observation of the two flare ribbons sandwiching, and the flaring loops striding, the mini-filament suggests that the magnetic arcade embedding the mini-filament has been torn into two parts, most likely through internal reconnection \cite[]{Gibson&Fan2006}. The upper part erupts with the blowout jet, and the lower part still holds the mini-filament in the low atmosphere, which is probably aided by the bald patch field (\S\ref{subsec:magnetic_config}). The internal reconnection might start early, as demonstrated by bright 131~{\AA} loops straddling the mini-filament during the standard-jet phase before the M1.2-class flare (Fig.~\ref{fig:flares}e). During the flare, the southern HXR footpoint source (25--45 keV) is apparently located below the fibrils involved in the blowout jet (Fig.~\ref{fig:halpha}e), which could be attributed to the interchange reconnection between the sheared arcade overlying the mini-filament and the fibrils. Thus, the mini-filament partially erupts in two aspects: 1) the filament itself exchanges mass and field with neighboring fibrils, and 2) only part of the magnetic arcade overlying the filament escapes with the jet.  

On the other hand, the transition from the standard to blowout jet clearly involves both the mass and fields of the fibrils located in the `upstream' of the single-strand spire of the standard jet, with these fibrils moving toward the spire. The fibril motions may trigger the blowout jet as suggested by the close temporal and spatial relation between them (Fig.~\ref{fig:slit}), and by the absence of obvious changes in other relevant features prior to the blowout jet, including the base arch and the single-strand spire. Particularly, the spiraling morphology and sweeping direction of these fibrils are consistent with the clockwise rotation of the sunspot.  Despite the difficulty of following the evolution of each individual fibrils, we  suggest that the fibril motions are an indication that the magnetic field associated with the fibrils changes orientation, which could be driven by the rotation of the sunspot. Such sweeping motions and changes of field orientations may make reconnections at the inverted Y-point of the single spire more favourable and therefore proceed faster, potentially triggering the eruption. In this regard, the sunspot rotation is not only crucial to the energy buildup but also to the trigger of the blowout jet.

%%%%%%%%%%%%%%%%%%%%%%%%%%%%%%%%%%%%%%%%%%%%%%%%%%%%%%%%%%%%%%%%%%%%%%%%%%%

\begin{authorcontribution}
T.G. and R.L. led the study, performed the analysis, and wrote the manuscript. Y.S. provided and analyzed the HXI data and contributed to the discussion. A.M.V. contributed to the interpretation and discussion. H.P. contributed to the STIX data analysis. R.B.L contributed to the NVST observation and data. W.G. is the PI of ASO-S mission and contributed to the discussion. All authors reviewed the manuscript.
\end{authorcontribution}

\begin{fundinginformation}
R.L. and Y.S. acknowledge the support from the National Key R\&D Program of China 2022YFF0503002. R.L., H.P., and R.B.L. acknowledge the support by NSFC (11925302, 42188101 and 42274204) and by the Strategic Priority Program of the Chinese Academy of Sciences (XDB41000000).
Y.S. and W.G. also acknowledge the support by NSFC (12333010, 11820101002, 11921003, 12233012). A.M.V acknowledges support by the Austrian Science Fund (FWF) 10.55776/I4555. 
\end{fundinginformation}

\begin{dataavailability}
The solar data used in the study are publicly available for download from the mission archives. The SDO data are available at \url{http://jsoc.stanford.edu/}. The NVST data are available at \url{https://fso.ynao.ac.cn/DataService/query}. The STIX data are available at \url{https://datacenter.stix.i4ds.net/}. The HXI data used in this study are during the mission commissioning phase and are available upon reasonable request, and the data after April 2023 are publicly available for download at \url{http://aso-s.pmo.ac.cn/sodc/dataArchive.jsp}.
\end{dataavailability}

\begin{ethics}
\begin{conflict}
The authors declare no conflicts of interest.
\end{conflict}
\end{ethics}

\bibliographystyle{spr-mp-sola}
\bibliography{nvst}

\end{document}